\documentclass[aps,prl,reprint,superscriptaddress,nobibnotes,floatfix,amsmath,amssymb,longbibliography]{revtex4-2}
\usepackage{graphicx,bm,booktabs,microtype}
\usepackage{tikz}
\usetikzlibrary{calc}
\usepackage[hidelinks]{hyperref}
\hypersetup{pdftitle={Anomalous Hall Response and Real-Space Current Correlations in Rhombohedral Graphene},pdfauthor={Zhen Bi and Chandra M. Varma}}
\newcommand{\dd}{\mathrm d}

\newcommand{\newsec}[1]{\textit{#1}}
\begin{document}
\raggedbottom
\clubpenalty=10000
\widowpenalty=10000
\brokenpenalty=10000
\title{Anomalous Hall Response and Real-Space Current Correlations in Rhombohedral Graphene}
\author{Zhen Bi}
\affiliation{Department of Physics, The Pennsylvania State University, University Park, Pennsylvania 16802, USA}
\affiliation{Center for Theory of Emergent Quantum Matter, Institute for Computational and Data Sciences, The Pennsylvania State University, University Park, Pennsylvania 16802, USA}
\author{Chandra M. Varma}
\thanks{Visiting Scholar at the University of California, Berkeley; Emeritus at the University of California, Riverside.}
\affiliation{Department of Physics, University of California, Berkeley, CA 94720}
\affiliation{Department of Physics, University of California, Riverside, CA 92521, USA}
\begin{abstract}
We investigate the anomalous Hall effect (AHE) and magnetization observed in rhombohedral graphene through real-space calculations which lead to new insights. Our calculations reveal no equilibrium loop currents of the type associated with the Haldane model of a quantum anomalous Hall insulator. Instead, the AHE arises in this instance from interference between electronic paths forming oriented closed Wilson loops of current operators in the presence of valley polarization. We show some examples of these Wilson loops. Using the fact that the added-carrier charge density resides dominantly in the top and bottom layers, we develop a real-space expansion of the surface Hamiltonian that reproduces the Hall response of the full model with good accuracy. We also predict that a smooth local potential induces equilibrium circulating currents whose strength depends on both Berry curvature and the orbital moments of states at the Fermi surface.
\end{abstract}
\maketitle

\newsec{Introduction} --- The discovery of the anomalous Hall effect (AHE) in metallic rhombohedral graphene is an exciting recent development in condensed matter physics~\cite{Zhou2021a,Liu2024,HanMultiferroic2023}. Related phenomena have been observed in graphene moiré superlattices~\cite{Choi2025,Lu2024}, spin--orbit-proximitized rhombohedral graphene~\cite{HanQAH2024}, and twisted MoTe$_2$~\cite{Cai2023,Park2023}. Recent nanoSQUID measurements on rhombohedral graphene directly image magnetic domains and reveal a large differential orbital moment in the electron-doped, spin--valley-polarized quarter metal~\cite{Sheekey2026}, while complementary measurements resolve isospin textures in tetralayer graphene~\cite{Auerbach2025}. Superconductivity occurs in nearby regions of density and displacement field~\cite{Zhou2021b}, and the observed spontaneous superconducting diode effect has been interpreted as evidence for chiral superconductivity~\cite{Han2025}. Recent nanoSQUID measurements also report time-reversal symmetry breaking in the superconducting state of pentalayer graphene~\cite{Dutta2026}. Understanding superconductivity of this kind requires understanding the symmetry and orbital character of the normal state~\cite{Varma2025diode,Chen2025}.

The theory of the AHE began with the work of Karplus and Luttinger on ferromagnets with spin--orbit coupling~\cite{Karplus1954,Luttinger1958}. In an inspired work, Haldane showed that complex transfer integrals breaking time reversal can support a quantized Hall response without net magnetic flux through the unit cell, generically accompanied by equilibrium circulating bond currents~\cite{Haldane1988}. Further-neighbor interactions can stabilize spontaneous loop-current order, with or without a topologically nontrivial band structure~\cite{Varma1997,Zhu2013,Raghu2008}. The intrinsic AHE is usually calculated in momentum space in terms of Berry curvature~\cite{Xiao2010,Nagaosa2010,Zhang2011}. Using this approach, Sheekey et al.~\cite{Sheekey2026} calculated the Hall response and orbital magnetization of the full eight-orbital tetralayer model with assumed spin and valley polarization. Disorder-induced contributions to the AHE in rhombohedral graphene have also been studied~\cite{Mikheeva2026}; here we consider the intrinsic response. The question we address here is whether the AHE in this state accompanies equilibrium loop currents or arises from interference between alternative propagation paths connecting the same initial and final sites in closed current correlations, while equilibrium bond currents vanish.

We investigate this question in real space by projecting the full Hamiltonian onto the top and bottom surface orbitals, where the low-energy states primarily reside. We find no equilibrium bond currents in the uniform valley-polarized state. The AHE instead comes from interference between electronic paths in closed correlations of current operators, which have a Wilson-loop structure. We illustrate the paths through the layers and compare the Hall response as additional transfer integrals are included. We then show that a smooth local potential produces equilibrium circulating currents and calculate their magnetic field.

\begingroup
\setlength{\abovedisplayskip}{6pt plus 2pt minus 1pt}
\setlength{\belowdisplayskip}{6pt plus 2pt minus 1pt}
\setlength{\abovedisplayshortskip}{3pt plus 1pt}
\setlength{\belowdisplayshortskip}{4pt plus 1pt}
\setlength{\jot}{2pt}
\setlength{\arraycolsep}{4pt}
\newsec{Full ABCA Hamiltonian} --- Sheekey et al.~\cite{Sheekey2026} use a continuum $\mathbf k\cdot\mathbf p$ Hamiltonian and also provide the tight-binding model needed for our real-space calculation. We consider ABCA tetralayer graphene with one $p_z$ orbital on each carbon atom. There are two sites, $A_\ell$ and $B_\ell$, in each layer $\ell$. Neglecting spin--orbit coupling, the Hamiltonian for one spin is
\begin{equation}
H=\sum_{i<j}\left(t_{ij}c_i^\dagger c_j+\mathrm{H.c.}\right)+\sum_i U_i n_i,
\label{eq:atomic_main}
\end{equation}
where the hopping amplitudes $t_{ij}$ are real. The largest, $t_0$, connects nearest neighbors within each layer, while $t_1$ connects the vertically aligned sites $B_\ell$ and $A_{\ell+1}$. The skew transfer $t_3$ connects $A_\ell$ and $B_{\ell+1}$, while $t_4$ connects $A_\ell,A_{\ell+1}$ and $B_\ell,B_{\ell+1}$. The transfer $t_2$ connects $A_\ell$ and $B_{\ell+2}$. The potential $U_i=U_\ell$ on layer $\ell$ accounts for the applied displacement field.

For valleys $\tau=\pm1$, write $\mathbf k=\tau\mathbf K+\mathbf q$ and define $\pi_\tau=\tau q_x+iq_y$, $q=|\mathbf q|$, and $v_i=\sqrt3at_i/(2\hbar)$ for $i=0,3,4$, where $a$ is the lattice constant. In the basis $(A_1,B_1,\ldots,A_4,B_4)$, the eight-band Hamiltonian $H_8$ has the following layer blocks to linear order in $\mathbf q$:
\begin{align}
H_{\ell\ell}&=\begin{pmatrix}
U_\ell&-\hbar v_0\pi_\tau\\
-\hbar v_0\pi_\tau^*&U_\ell
\end{pmatrix},\label{eq:model}\\
H_{\ell,\ell+1}&=\begin{pmatrix}
-\hbar v_4\pi_\tau&-\hbar v_3\pi_\tau^*\\
t_1&-\hbar v_4\pi_\tau
\end{pmatrix},\quad H_{\ell,\ell+2}=\begin{pmatrix}
0&t_2\\
0&0
\end{pmatrix}.\label{eq:interlayer_blocks}
\end{align}
Hermiticity supplies the lower blocks. The signs and complex conjugates follow from the bond vectors given in End Matter. In the full lattice model, $-\hbar v_i\pi_\tau$ is replaced by $t_if(\mathbf k)$, with $f(\mathbf k)=\sum_{j=1}^3e^{i\mathbf k\cdot\boldsymbol\delta_j}$. We use this lattice form for the real-space paths and numerical calculations.

\begin{figure*}[t]
\centering
\begin{tikzpicture}[x=1cm,y=1cm,every node/.style={font=\scriptsize},site/.style={circle,fill=black,inner sep=1.3pt},routea/.style={blue!65!black,line width=.85pt},routeb/.style={orange!85!black,dashed,line width=.85pt}]
\begin{scope}[shift={(0.1,0)}]
\node[anchor=west,font=\small] at (-1.6,.65) {(a) $t_0,t_1$};
\coordinate (a0) at (0,0);
\coordinate (a1) at (-1.05,-.55);
\coordinate (a2) at (-1.05,-1.10);
\coordinate (a3) at (-1.05,-1.65);
\coordinate (a4) at (-1.05,-2.20);
\coordinate (a5) at (-1.05,-2.75);
\coordinate (a6) at (-1.05,-3.30);
\coordinate (a7) at (0,-3.85);
\coordinate (a8) at (1.05,-3.30);
\coordinate (a9) at (1.05,-2.75);
\coordinate (a10) at (1.05,-2.20);
\coordinate (a11) at (1.05,-1.65);
\coordinate (a12) at (1.05,-1.10);
\coordinate (a13) at (1.05,-.55);
\foreach \i/\j/\t in {0/1/0,1/2/1,2/3/0,3/4/1,4/5/0,5/6/1,6/7/0}{
\draw[routea] (a\i)--node[right,font=\tiny,fill=white,inner sep=.3pt]{$t_{\t}$}(a\j);}
\draw[routeb] (a0)--(a13)--(a12)--(a11)--(a10)--(a9)--(a8)--(a7);
\node[site,label=above:{$A_1$}] at(a0){};
\node[site,label=below:{$B_4$}] at(a7){};
\foreach \i/\lab in {1/B_1,2/A_2,3/B_2,4/A_3,5/B_3,6/A_4}{\node[site,label=left:{$\lab$}] at(a\i){};}
\foreach \i/\lab in {8/A_4,9/B_3,10/A_3,11/B_2,12/A_2,13/B_1}{\node[site,label=right:{$\lab$}] at(a\i){};}
\end{scope}
\begin{scope}[shift={(4.5,0)}]
\node[anchor=west,font=\small] at (-1.6,.65) {(b) $t_0,t_1,t_3$};
\coordinate (n0) at (0.000,0.000);\node[site,label=above:{$A_{1}$}] at(n0){};
\coordinate (n1) at (-1.050,-1.200);\node[site,label=left:{$B_{1}$}] at(n1){};
\coordinate (n2) at (-1.050,-2.400);\node[site,label=left:{$A_{2}$}] at(n2){};
\coordinate (n3) at (0.000,-3.600);\node[site,label=below:{$B_{2}$}] at(n3){};
\draw[routea] (n0)--node[right,font=\tiny,fill=white,inner sep=.5pt]{$t_0$}(n1);
\draw[routea] (n1)--node[right,font=\tiny,fill=white,inner sep=.5pt]{$t_1$}(n2);
\draw[routea] (n2)--node[right,font=\tiny,fill=white,inner sep=.5pt]{$t_0$}(n3);
\draw[routeb] (n0)--node[left,font=\tiny,fill=white,inner sep=.5pt]{$t_3$}(n3);
\end{scope}
\begin{scope}[shift={(8.2,0)}]
\node[anchor=west,font=\small] at (-1.6,.65) {(c) $t_0,t_1,t_4$};
\coordinate (n0) at (0.000,0.000);\node[site,label=above:{$A_1(i)$}] at(n0){};
\coordinate (n1) at (-1.050,-1.200);\node[site,label=left:{$B_1$}] at(n1){};
\coordinate (n2) at (-1.050,-2.400);\node[site,label=left:{$A_2$}] at(n2){};
\coordinate (n3) at (0.000,-3.600);\node[site,label=below:{$A_1(j)$}] at(n3){};
\coordinate (n4) at (1.050,-2.400);\node[site,label=right:{$B_1$}] at(n4){};
\coordinate (n5) at (1.050,-1.200);\node[site,label=right:{$A_2$}] at(n5){};
\draw[routea] (n0)--node[right,font=\tiny,fill=white,inner sep=.5pt]{$t_0$}(n1);
\draw[routea] (n1)--node[right,font=\tiny,fill=white,inner sep=.5pt]{$t_1$}(n2);
\draw[routea] (n2)--node[right,font=\tiny,fill=white,inner sep=.5pt]{$t_4$}(n3);
\draw[routeb] (n0)--node[left,font=\tiny,fill=white,inner sep=.5pt]{$t_4$}(n5);
\draw[routeb] (n5)--node[left,font=\tiny,fill=white,inner sep=.5pt]{$t_1$}(n4);
\draw[routeb] (n4)--node[left,font=\tiny,fill=white,inner sep=.5pt]{$t_0$}(n3);
\end{scope}
\begin{scope}[shift={(12.6,0)}]
\node[anchor=west,font=\small] at (-1.6,.65) {(d) Joining small loops};
\coordinate (b0) at (0,0);
\coordinate (b1) at (-.95,-.55);
\coordinate (b2) at (-.95,-1.10);
\coordinate (b3) at (.95,-1.65);
\coordinate (b4) at (.95,-2.20);
\coordinate (b5) at (-.95,-2.75);
\coordinate (b6) at (-.95,-3.30);
\coordinate (b7) at (0,-3.85);
\fill[blue!8] (b0)--(b1)--(b2)--(b3)--cycle;
\fill[orange!12] (b2)--(b3)--(b4)--(b5)--cycle;
\fill[teal!9] (b4)--(b5)--(b6)--(b7)--cycle;
\draw[gray!70,densely dashed,line width=.7pt] (b2)--node[above,sloped,font=\tiny]{$t_0$}(b3);
\draw[gray!70,densely dashed,line width=.7pt] (b4)--node[above,sloped,font=\tiny]{$t_0$}(b5);
\foreach \i/\j/\t in {0/1/0,1/2/1,2/5/3,5/6/1,6/7/0}{\draw[->,line width=.85pt] (b\i)--node[right,font=\tiny,fill=white,inner sep=.3pt]{$t_{\t}$}(b\j);}
\foreach \i/\j/\t in {7/4/3,4/3/1,3/0/3}{\draw[->,line width=.85pt] (b\i)--node[left,font=\tiny,fill=white,inner sep=.3pt]{$t_{\t}$}(b\j);}
\node[blue!65!black] at (-.17,-.72) {$L_{12}$};
\node[orange!75!black] at (0,-1.93) {$-L_{23}$};
\node[teal!75!black] at (-.17,-3.13) {$L_{34}$};
\node[site,label=above:{$A_1$}] at(b0){};
\node[site,label=below:{$B_4$}] at(b7){};
\foreach \i/\lab in {1/B_1,2/A_2,5/B_3,6/A_4}{\node[site,label=left:{$\lab$}] at(b\i){};}
\foreach \i/\lab in {3/B_2,4/A_3}{\node[site,label=right:{$\lab$}] at(b\i){};}
\node at (0,-4.43) {$L=L_{12}-L_{23}+L_{34}$};
\end{scope}
\end{tikzpicture}
\caption{Illustrative loops and their composition. (a) Two $01$ routes form a four-layer contour. (b) A $t_3$ bond replaces a $t_0,t_1,t_0$ route. (c) Two $t_0,t_1,t_4$ routes connect sites on the same surface; the shared dimer is drawn twice. (d) Shared dashed bonds cancel, leaving the black outer contour. Arrows mark path orientation. Connectivity is schematic. Hall weights are not additive under contour composition.}
\label{fig:surface_series}
\end{figure*}
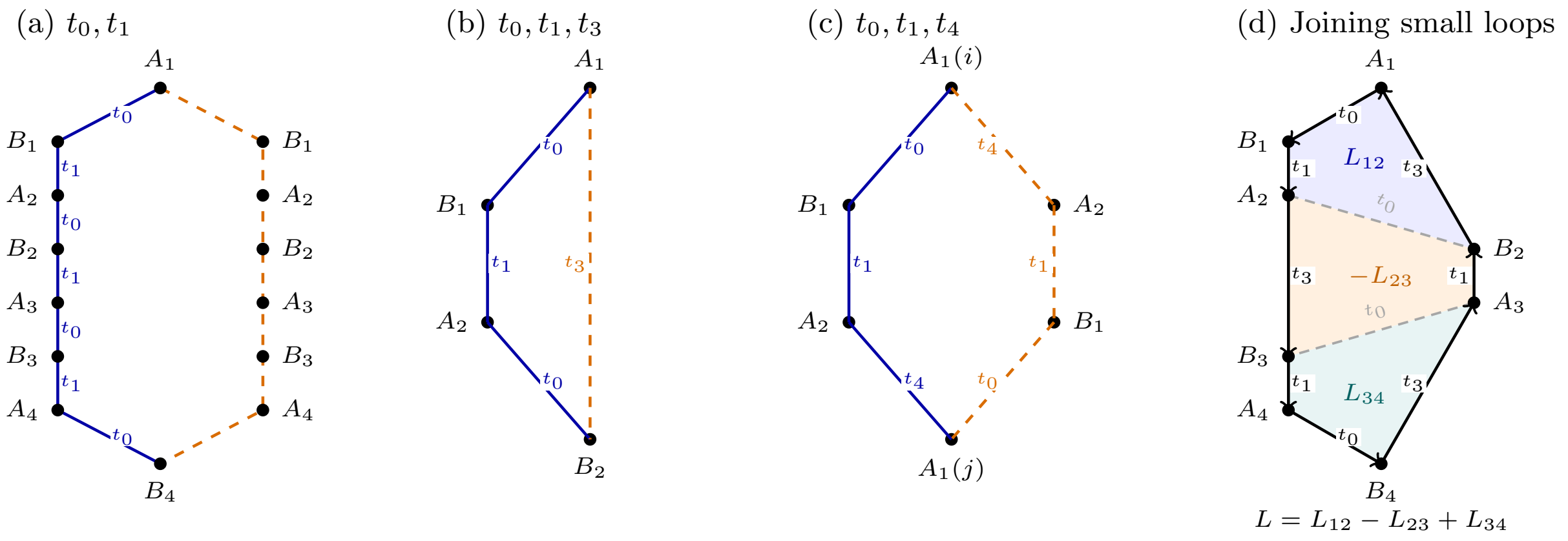

\newsec{Reduction to the two outer surfaces} --- The low-energy states of the model lie mostly on the top and bottom layers~\cite{Koshino2009,Zhang2010}. For the density and layer potentials used below, approximately $92.85\%$ of the added-carrier charge resides in these two layers. It is worthwhile to understand the microscopic physics by reducing the model to a simpler one which retains the phenomena. 

Let $P$ project the Hamiltonian onto $(A_1,B_4)$, i.e., the $A$ sites in layer 1 and the $B$ sites in layer 4, and let $Q=1-P$. Eliminating the interior orbitals gives the exact energy-dependent surface Hamiltonian~\cite{Lowdin1951}
\begin{equation}
H_{\rm eff}(E,\mathbf q)=H_{PP}+H_{PQ}(E-H_{QQ})^{-1}H_{QP}.
\label{eq:surface_projection}
\end{equation}
This projection applies to the full model and to each reduced hopping model. The convergence of an expansion of the interior propagator is tested in the Supplemental Material~\cite{SM}. The surface Hamiltonians below refer to a single valley, with $\mathbf q$ measured from its center.

Let us first put $t_2=t_3=t_4=0$. We call this the $01$ model. Keeping the surface potentials $U_1=-U_4=\Delta$ and the leading transfer between the surfaces gives
\begin{equation}
H_{01}(\mathbf q)\simeq
\begin{pmatrix}
\Delta&-\alpha_4\pi_\tau^4\\
-\alpha_4(\pi_\tau^*)^4&-\Delta
\end{pmatrix},\qquad
\alpha_4=\frac{(\hbar v_0)^4}{t_1^3}.
\label{eq:h01_surface}
\end{equation}
Here $2\Delta$ is the potential difference between the surfaces. This simple form omits the energy dependence and momentum-dependent diagonal corrections of the exact projection (End Matter). The off-diagonal phase winds four times around a valley. At $\Delta=0$, the low-energy dispersion is proportional to $q^4$, as in the full eight-orbital $01$ model~\cite{Koshino2009}. Although $\Delta<t_1$, it distinguishes the surfaces and changes the band-edge states. The expansion parameter is $\lambda_q=\hbar v_0q/t_1$, not $t_0/t_1$.

Adding $t_3$ gives the $013$ model. For this and the $014$ model below, we evaluate the virtual transfers at $E=U_\ell=0$, retaining $\pm\Delta$ on the surfaces. The correction $H_{013}=H_{01}+\delta H_3$ is
\begin{equation}
\begin{aligned}
\delta H_3&=\begin{pmatrix}0&\delta h_3\\\delta h_3^*&0\end{pmatrix},\\
\delta h_3&=-3\frac{(\hbar v_0\pi_\tau)^2(\hbar v_3\pi_\tau^*)}{t_1^2}
-\frac{(\hbar v_3\pi_\tau^*)^2}{t_1}.
\end{aligned}
\label{eq:h013}
\end{equation}
These additional transfers change the magnitude and phase of the coupling between the surfaces and produce trigonal warping~\cite{Koshino2009}. 

Adding $t_4$ instead gives the $014$ model. Virtual motion through a vertical dimer produces the same leading energy shift on both surfaces, $H_{014}=H_{01}+\delta H_4$, with
\begin{equation}
\begin{aligned}
\delta H_4&=\begin{pmatrix}d_4&0\\0&d_4\end{pmatrix},\\
d_4&=-\frac{2t_0t_4}{t_1}|f|^2\simeq-\frac{2\hbar^2v_0v_4}{t_1}q^2.
\end{aligned}
\label{eq:d014}
\end{equation}
This leading term is linear in $t_4$. For our sign of $t_4$, it raises the energy away from the valley center and can change the occupied region at fixed density, while leaving the surface eigenstates unchanged at this order. The derivation and corrections from the energy dependence and layer potentials are given in End Matter.

We assume that interactions polarize the added electrons into one spin and one valley, as in the quarter-metal state. At zero temperature, the added electrons fill the conduction states of this channel in order of energy up to the prescribed carrier density, while the valence bands remain filled in both spins and valleys. We use the bare tight-binding bands throughout.

\endgroup

\newsec{No equilibrium currents in the uniform state} --- The current on a bond is proportional to $\operatorname{Im}[t_{ij}\langle c_i^\dagger c_j\rangle]$. For real transfer integrals, equal occupation of time-reversed states gives zero equilibrium bond current. Valley polarization removes this equality. In the uniform state considered here, the bond currents nevertheless vanish because of the combined reflection and time-reversal symmetry.

To see this, choose $x$ along the line joining the valleys and consider $M_x:(x,y,z)\mapsto(-x,y,z)$. Reflection and orbital time reversal $\mathcal T_0$, which complex conjugates the orbital wave function, each exchange the valleys, but their product preserves the polarized state. A bond whose endpoints lie in the mirror plane is fixed by reflection and has its current reversed by time reversal. Its current must therefore vanish. Translations and threefold rotations cover all $t_0$--$t_4$ bonds in this way. The Hall response and perpendicular magnetization are unchanged under the combined operation and remain allowed. 

\newsec{Hall response as oriented real-space loops} --- The intrinsic Hall response is determined by the correlation of two current operators~\cite{Kubo1957},
\begin{equation}
\begin{aligned}
\sigma_{xy}&=-\lim_{\omega\to0}\frac{\Pi^R_{xy}(\omega)-\Pi^R_{xy}(0)}{i\omega},\\
\Pi^R_{xy}(t)&=-\frac{i}{\hbar A}\theta(t)\langle[\widehat J_x(t),\widehat J_y]\rangle,
\end{aligned}
\label{eq:kubo_main}
\end{equation}
where $\widehat J_a$ is the total current, $A$ is the sample area, and the antisymmetric $xy$ part is taken. We use $\Pi^R_{xy}(\omega)=\int_0^\infty\dd t\,e^{i(\omega+i0^+)t}\Pi^R_{xy}(t)$, with the perturbation $-\widehat{\mathbf J}\cdot\mathbf A(t)$ and electric field $\mathbf E=-\partial_t\mathbf A$. Time evolution is generated by the bare Hamiltonian, and the expectation value is taken in the prescribed zero-temperature valley-polarized state.

For the quadratic Hamiltonian, contracting the fermion operators gives
\begin{equation}
\Pi^R_{xy}(t)=\frac{2\theta(t)}{\hbar A}\operatorname{Im}\operatorname{Tr}[J_xG_+J_yG_-],
\label{eq:green_main}
\end{equation}
where $G_+=G^{\rm T}(t)$ and $G_-=G^{\rm T}(-t)$ are the time-ordered Green functions of the prescribed state. Here $\widehat J_a=c^\dagger J_ac$, with $J_a$ the physical current matrix of the bare hopping Hamiltonian. The Green functions contain the bare-band energies and the chosen occupations (End Matter).

In the atomic basis, the trace contains the closed products
\begin{equation}
(J_x)_{ij}(G_+)_{jk}(J_y)_{kl}(G_-)_{li},
\label{eq:green_loop}
\end{equation}
with all indices summed. The two Green functions connect the current bonds and close the path. On a current bond, the transfer integral is replaced by $(J_a)_{ij}=-ie(r_j-r_i)_aH_{ij}/\hbar$, where $e>0$. The closed products are invariant under changes of atomic-orbital phases. These are the Wilson loops of the current correlation. Their Hall contribution follows from antisymmetrizing the current directions and taking the Kubo time integral.

The bare hopping product around any contour in Fig.~\ref{fig:surface_series} is real; for example, the four-bond contour in panel (b) gives $t_0^2t_1t_3$. The Hall response also depends on the propagation and occupation weights. With equal valley occupations, the Green functions are symmetric and $J_a^{\mathsf T}=-J_a$ in the real atomic basis. Transposing Eq.~\eqref{eq:green_main} gives $\Pi_{xy}=\Pi_{yx}$, so the Hall response vanishes. Valley polarization changes the propagators and removes this cancellation. These correlations can give a Hall response without equilibrium bond currents. 

The Hall response comes from interference between propagation paths. Schematically, in momentum space, a path has amplitude $\mathcal A_\gamma(\mathbf k)=a_\gamma e^{i\mathbf k\cdot\mathbf R_\gamma}$, where $a_\gamma$ is momentum independent and $\mathbf R_\gamma$ is its net displacement. Two paths give the cross term $a_\gamma a_{\gamma'}^*e^{i\mathbf k\cdot(\mathbf R_\gamma-\mathbf R_{\gamma'})}$. Momentum derivatives in the current operators give an antisymmetric interference term proportional to $R_{\gamma x}R_{\gamma' y}-R_{\gamma y}R_{\gamma' x}$. This is the real-space content of the imaginary, antisymmetric product entering the Berry curvature. Opposite valleys have opposite curvatures~\cite{XiaoYaoNiu2007},
\begin{equation}
\Omega_{n,z}(\mathbf K+\mathbf q)=-\Omega_{n,z}(\mathbf K'-\mathbf q).
\label{eq:valley_curvature}
\end{equation}
Valley polarization gives unequal occupations, $n_K\ne n_{K'}$, so the two valley contributions need not cancel.

\newsec{Real-space paths and the Hall response} --- The paths in Fig.~\ref{fig:surface_series} illustrate the virtual transfers underlying the terms in the surface Hamiltonian. We use the $01$, $013$ and $014$ models to examine the paths generated by the relevant hopping process and compare their Hall responses with the full model. For each model, we diagonalize the corresponding eight-orbital Hamiltonian and calculate its current matrices from $J_a=-(e/\hbar)\partial_{k_a}H_8$. All eight bands enter the intermediate-state sum.

\textit{Paths in the $t_0,t_1$ model.} A monotonic path through the stack is
\begin{equation}
\begin{aligned}
A_1&\xrightarrow{t_0}B_1\xrightarrow{t_1}A_2\xrightarrow{t_0}B_2\\
&\xrightarrow{t_1}A_3\xrightarrow{t_0}B_3\xrightarrow{t_1}A_4\xrightarrow{t_0}B_4.
\end{aligned}
\label{eq:path01}
\end{equation}
At each $t_0$ step there are three nearest-neighbor bond directions, while $t_1$ connects vertically aligned sites. Figure~\ref{fig:surface_series}(a) shows two routes between the same surface sites. Following one forward and the other backward forms a closed contour.

\textit{Additional paths in the $t_0,t_1,t_3$ model.} All $01$ paths remain present. In addition, $t_3$ closes the elementary four-bond contour
\begin{equation}
A_\ell\xrightarrow{t_0}B_\ell\xrightarrow{t_1}A_{\ell+1}\xrightarrow{t_0}B_{\ell+1}\xrightarrow{t_3}A_\ell.
\label{eq:path013}
\end{equation}
The $t_3$ bond replaces the three-step $t_0,t_1,t_0$ route [Fig.~\ref{fig:surface_series}(b)]. There are three places for one replacement along Eq.~\eqref{eq:path01}, and one way to make two, giving the two terms in $\delta h_3$, Eq.~\eqref{eq:h013}.

\textit{Additional paths in the $t_0,t_1,t_4$ model.} In Fig.~\ref{fig:surface_series}(c), an electron enters a vertical dimer through $t_0$ and returns to the same surface through $t_4$, or takes these steps in the opposite order. Summing these transfers gives the energy correction in Eq.~\eqref{eq:d014}. The paths thus affect the Hall response through changes in energy and occupation as well as interference.

Translations and threefold rotations generate equivalent loops. Joining neighboring loops with opposite orientations on their shared bonds gives larger contours, as in Fig.~\ref{fig:surface_series}(d), where $L=L_{12}-L_{23}+L_{34}$. This is a relation between paths; their Hall weights also depend on propagation and occupation. The loops shown are illustrative; the Hall calculation includes all paths of each model and their interference.

Table~\ref{tab:model_hall} compares the Hall conductivities at carrier density $n_c=4.695\times10^{11}$ cm$^{-2}$ and layer potentials $U_\ell=(25,25/3,-25/3,-25)$ meV. We use $a=0.246$ nm, $(t_0,t_1)=(2.6,0.3561)$ eV, and $(t_2,t_3,t_4)=(-0.0083,-0.293,-0.144)$ eV, with the occupied states determined for each Hamiltonian. The $01$ and $013$ models overestimate the magnitude of the response. Including $t_4$ brings the $014$ result within $0.19\%$ of the full model; the $0134$ error is $2.68\%$. At these parameters, $014$ gives the closest agreement among the reduced models considered.

\begin{table}[htbp]
\caption{Zero-temperature intrinsic Hall conductivity at the density and layer potentials specified in the text. Each model uses only the listed hoppings, with its own bands, occupations, and current operators. Relative error: $|\sigma_{xy}-\sigma_{xy}^{\rm full}|/|\sigma_{xy}^{\rm full}|$.}
\label{tab:model_hall}
\begin{ruledtabular}
\begin{tabular}{lrr}
Hoppings retained & $\sigma_{xy}/(e^2/h)$ & Relative error\\
$t_0,t_1$ & $-0.39567$ & $71.37\%$\\
$t_0,t_1,t_3$ & $-0.43316$ & $87.61\%$\\
$t_0,t_1,t_4$ & $-0.23132$ & $0.19\%$\\
$t_0,t_1,t_2,t_3$ & $-0.40391$ & $74.94\%$\\
$t_0,t_1,t_3,t_4$ & $-0.22470$ & $2.68\%$\\
$t_0,t_1,t_2,t_3,t_4$ & $-0.23089$ & $0.00\%$\\
\end{tabular}
\end{ruledtabular}
\end{table}

\begin{figure}[!htb]
\centering
\includegraphics[width=0.96\columnwidth]{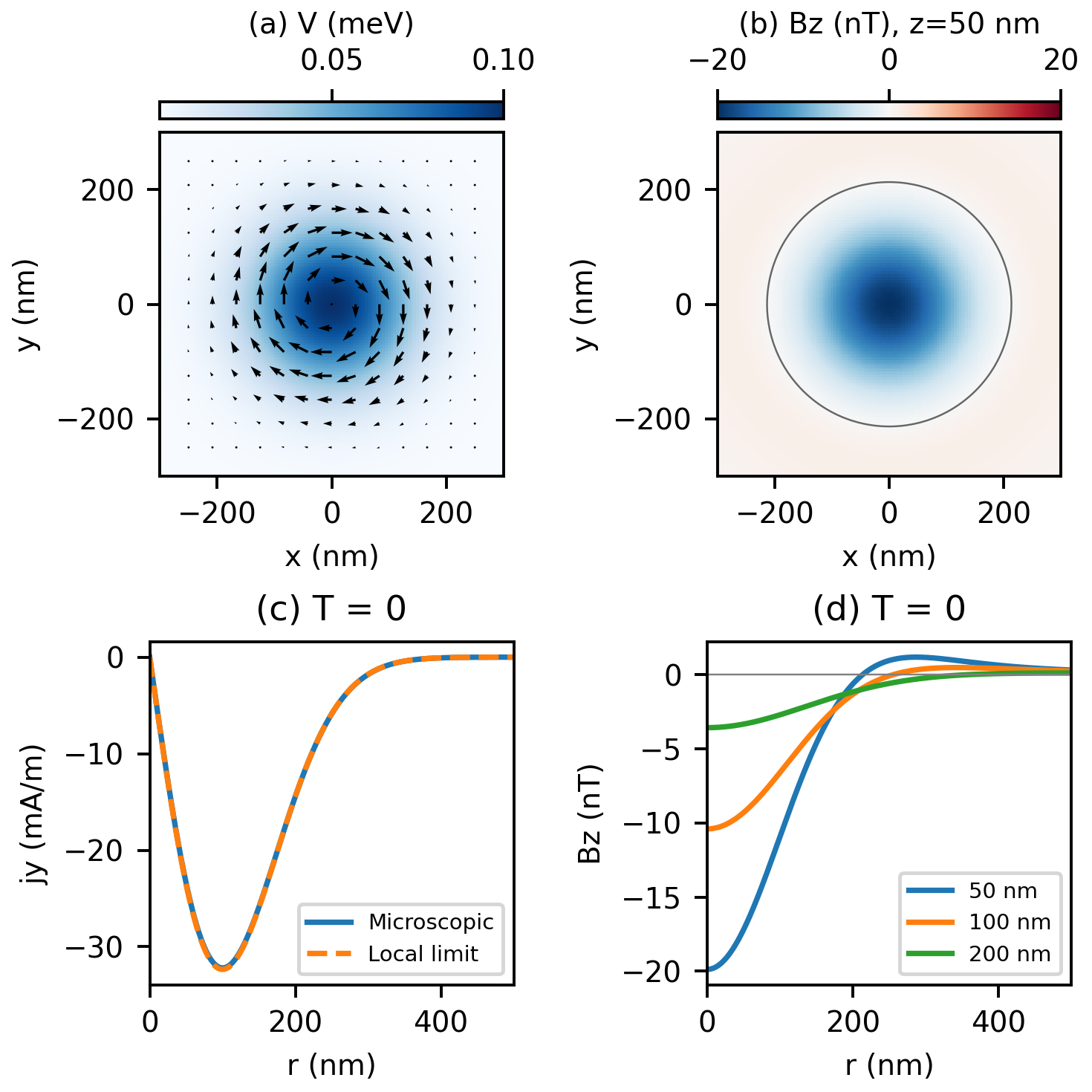}
\caption{Zero-temperature current and magnetic field induced by a Gaussian potential of height $0.1$ meV and width $100$ nm in the $+K$ state. (a) Charge-current arrows over the potential. (b) $B_z$ at height $50$ nm; the contour marks $B_z=0$. (c) Current along $(r,0)$: microscopic response and local approximation. (d) Magnetic-field profiles at three heights. Reversing valley reverses both current and field.}
\label{fig:defect_main}
\end{figure}

\newsec{Circulating currents around a defect} --- A screened charged-impurity potential~\cite{SMAdam} or a local gate can produce equilibrium circulating currents in the valley-polarized state. A smooth potential changes the local carrier density and orbital magnetization, giving rise to currents around the defect. Potential-induced equilibrium currents have been studied in gapped graphene and other chiral systems~\cite{Koshino2011,Mishchenko2014}. We calculate the response to a Gaussian potential $V(\mathbf r)=V_0e^{-r^2/(2\xi^2)}$ to first order in $V$. We take $V$ to be the screened potential and keep the valley polarization fixed, without a self-consistent Hartree or exchange correction.

For a slowly varying potential, the current is determined by the spatial variation of orbital magnetization~\cite{Cooper1997,XiaoYao2006}, and its strength is governed by $\partial M_z/\partial\mu$. This derivative contains a Fermi-surface orbital-moment term as well as a Berry-curvature term~\cite{Xiao2010,XiaoNiu2020}. Figure~\ref{fig:defect_main} shows that the microscopic current calculation agrees closely with this local description. At the parameters studied here, the orbital-moment term is larger. The circulation therefore probes orbital information beyond the Hall conductivity. Related calculations for rhombohedral trilayer graphene also find that the orbital magnetization can change sign while the spin and valley polarization remain unchanged~\cite{DasHuang2024}.

The resulting magnetic field changes sign away from the center of the defect. Reversing the occupied valley reverses both the current and the field. A modulated local gate could distinguish the induced magnetic signal from a static background. Electrostatic perturbations have already been used to induce and image equilibrium currents in graphene in the quantum Hall regime~\cite{Uri2020}. Comparison with nanoSQUID measurements~\cite{SMTschirhart} requires the magnetic flux through the pickup loop at the probe height.

\newsec{Discussion} --- The real-space calculation distinguishes two kinds of closed paths. The Hall effect involves correlations of current operators and interference between propagation paths. Equilibrium loop currents would instead require a nonzero current expectation value on the physical bonds. The uniform valley-polarized model has the first and not the second. The $01$, $013$ and $014$ models illustrate how these paths change the surface transfer, energies and occupations, and hence the Hall response. A smooth local potential produces equilibrium circulation, whose strength depends on the orbital moments as well as the Hall response.

Experiments in thicker rhombohedral samples have also revealed coupled in-plane and out-of-plane orbital responses in a state that breaks rotational symmetry~\cite{Li2026Transdimensional}. Extending the present analysis to that state would require allowing the additional symmetry breaking.

We have assumed complete spin and valley polarization of the added carriers. How the polarization changes near a defect or a domain boundary remains to be determined self-consistently. The nearby superconducting state raises a further question: how does pairing change these current correlations and the equilibrium currents near a defect? Their magnetic fields may help distinguish the superconducting orbital response from that already present in the polarized normal state.

\newsec{Acknowledgments} --- C.M.V. thanks Nadav Auerbach and David Goldhaber-Gordon for discussions of the experiments, and Tomo Soejima and Pavel Volkov for discussions of theoretical issues. Z.B. also acknowledges support from NSF under award number DMR-2339319.  This collaboration was initiated in part at the Aspen Center for Physics, which is supported by National Science Foundation grant PHY-2210452. We acknowledge the use of generative AI tools for numerical calculation and manuscript editing. 

\makeatletter
\immediate\write\@auxout{\string\citation{v4BibControl}}
\makeatother
\bibliographystyle{apsrev4-2}
\bibliography{graphene-loop}
\begingroup
\setlength{\dbltextfloatsep}{10pt plus 2pt minus 2pt}
\setlength{\abovedisplayskip}{6pt plus 2pt minus 1pt}
\setlength{\belowdisplayskip}{6pt plus 2pt minus 1pt}
\setlength{\abovedisplayshortskip}{3pt plus 1pt}
\setlength{\belowdisplayshortskip}{4pt plus 1pt minus 1pt}
\section*{End Matter}
\newsec{Model parameters and occupied state} --- In the atomic Bloch basis, the lattice Hamiltonian has blocks
\begin{equation*}
\begin{aligned}
H_{\ell\ell}&=\begin{pmatrix}U_\ell&t_0f\\t_0f^*&U_\ell\end{pmatrix},\\
H_{\ell,\ell+1}&=\begin{pmatrix}t_4f&t_3f^*\\t_1&t_4f\end{pmatrix},\qquad
H_{\ell,\ell+2}=\begin{pmatrix}0&t_2\\0&0\end{pmatrix}.
\end{aligned}
\end{equation*}
Here $f=\sum_\nu e^{i\mathbf k\cdot\boldsymbol\delta_\nu}$, with $\boldsymbol\delta_1=(0,-a/\sqrt3)$ and $\boldsymbol\delta_{2,3}=(\pm a/2,a/(2\sqrt3))$. We use $a=0.246$ nm, $(t_0,t_1,t_2,t_3,t_4)=(2.6,0.3561,-0.0083,-0.293,-0.144)$ eV, $U_\ell=(25,25/3,-25/3,-25)$ meV, and added density $n_c=4.695\times10^{11}$ cm$^{-2}$. The valleys lie at $\pm\mathbf K$, with $\mathbf K=(4\pi/3a,0)$ and $f(\mathbf K+\mathbf q)=-\sqrt3a(q_x+iq_y)/2+O(q^2)$.

The valence bands are filled, and the added electrons occupy one spin and valley. Their occupation is $\delta f_{n\mathbf k}=\Theta(\mu-\epsilon_{n\mathbf k})$ in the selected conduction channel and zero in the others, with $\mu$ fixed by the carrier density.

Valley polarization preserves lattice translations but produces complex intersite correlations. For sites $i=(\mathbf R,\alpha)$ and $j=(\mathbf R',\beta)$, the added carriers give
\begin{equation}
\delta\langle c_j^\dagger c_i\rangle=\frac1N\sum_{n\mathbf k}\delta f_{n\mathbf k}e^{i\mathbf k\cdot(\mathbf r_i-\mathbf r_j)}u_{n\mathbf k,\alpha}u_{n\mathbf k,\beta}^*,
\label{eq:occupation}
\end{equation}
where $N$ is the number of unit cells and $u_{n\mathbf k}$ is the Bloch eigenvector. Writing $\mathbf k=\mathbf K+\mathbf q$ displays the factor $e^{i\mathbf K\cdot(\mathbf r_i-\mathbf r_j)}$ multiplying the envelope from the occupied states near that valley. Reversing the valley conjugates the correlation. This pattern does not break translation symmetry or require intervalley coherence; its physical bond currents vanish by the symmetry discussed in the main text.

The intrinsic carrier Hall conductivity is~\cite{Xiao2010,Nagaosa2010}
\begin{equation}
\frac{\sigma_{xy}}{e^2/h}=-2\pi\sum_n\int\frac{\dd^2k}{(2\pi)^2}\delta f_{n\mathbf k}\Omega_{n\mathbf k}.
\label{eq:berry}
\end{equation}
The reference valence bands cancel between valleys. Keeping all eight intermediate bands gives $\sigma_{xy}=-0.23089\,e^2/h$ for the added electrons in $+\mathbf K$; the opposite valley reverses its sign.

\begin{figure*}[t]
\centering
\includegraphics[width=0.88\textwidth]{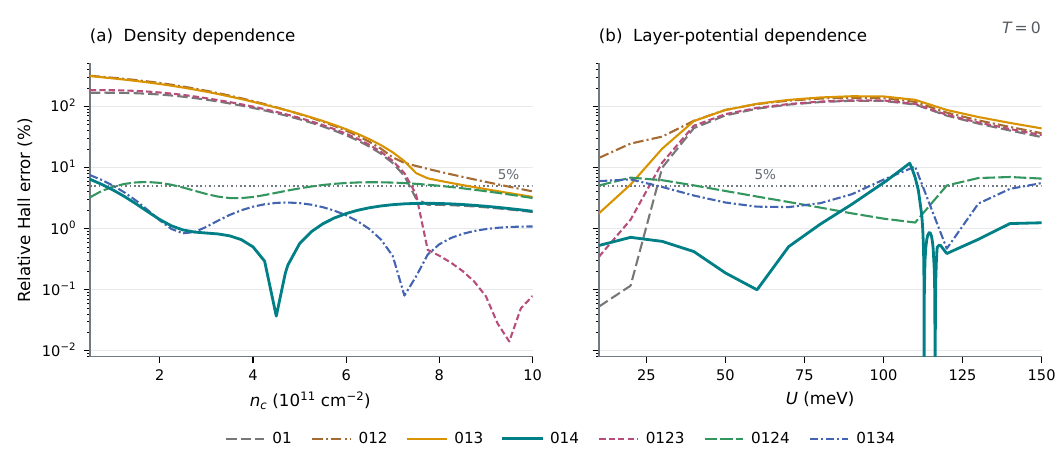}
\caption{Relative Hall-conductivity error, $|\sigma_{xy}-\sigma_{xy}^{\rm full}|/|\sigma_{xy}^{\rm full}|$, for reduced hopping models retaining $t_0,t_1$. (a) Density dependence at $U=50$ meV. (b) Layer-potential dependence at $n_c=4.695\times10^{11}$ cm$^{-2}$. The dotted line marks $5\%$. The two deep dips of the $014$ curve beyond its peak correspond to zero crossings of the signed error.}
\label{fig:model_parameter_dependence}
\end{figure*}

\newsec{Real-space expansion of the Hall response} --- First consider the unpolarized state. Write $H=\varepsilon+T$, with onsite energies $\varepsilon_i$ and transfer integrals $T_{ij}=t_{ij}$ for $i\ne j$. The resolvent has the formal hopping expansion
\begin{equation}
G=g+gTg+gTgTg+\cdots,\qquad g_i(z)=(z-\varepsilon_i)^{-1}.
\label{eq:onsite_expansion}
\end{equation}
Using this expansion for the frequency-domain propagators in Eq.~\eqref{eq:green_main} gives closed paths with two current insertions. A returning path $\gamma:i_1\to\cdots\to i_p\to i_1$ contains
\begin{equation}
W_\gamma=t_{i_1i_2}g_{i_2}(z_2)\cdots t_{i_pi_1}g_{i_1}(z_1),
\label{eq:onsite_loop}
\end{equation}
Here $z_r$ is the energy argument of the propagation segment containing site $i_r$. The current vertices supply the bond displacements. Pairing reversed paths gives an antisymmetric contribution proportional to $\operatorname{Im}\mathcal W_\gamma$, with $\mathcal W_\gamma=t_{i_1i_2}\cdots t_{i_pi_1}$. For real transfer integrals this vanishes.

For the prescribed polarized state, the bare eigenstates $|\nu\rangle$ have energies $\epsilon_\nu$ and occupations $n_\nu=0$ or $1$. The time-ordered Green function, $G^{\rm T}_{ij}(t)=-i\langle\mathcal T c_i(t)c_j^\dagger(0)\rangle$, is
\begin{equation}
G^{\rm T}(E)=\sum_\nu|\nu\rangle\langle\nu|\left[\frac{1-n_\nu}{E-\epsilon_\nu+i0^+}+\frac{n_\nu}{E-\epsilon_\nu-i0^+}\right].
\label{eq:valley_resolvent}
\end{equation}
Equal-energy states in opposite valleys can now have different occupations. These enter through the pole prescriptions and must be retained when expanding the propagators. The bare hopping product $\mathcal W_\gamma$ remains real, but the occupied-state projectors need not be real in the atomic basis. Reversing the occupied valley conjugates them and reverses the Hall response.

To connect the real-space current correlations with the band calculation, we write the dc response in spectral form~\cite{Garcia2015}. Define $\rho(E)=\delta(E-H)$ and the occupied spectral density $\rho_{\rm occ}(E)=\sum_\nu n_\nu|\nu\rangle\langle\nu|\delta(E-\epsilon_\nu)$. The latter retains the different occupations of equal-energy states in opposite valleys. With $J_a=-eD_a/\hbar$ and $(D_a)_{ij}=i(r_j-r_i)_aH_{ij}$, the response is
\begin{equation}
\begin{aligned}
\frac{\sigma_{xy}^{(\eta)}}{e^2/h}
&=\frac{2\pi}{A}\int\frac{\dd E\,\dd E'}{(E-E')^2+\eta^2}\\
&\quad\times\operatorname{Im}\operatorname{Tr}\bigl\{\rho_{\rm occ}(E)
[D_x\rho(E')D_y-D_y\rho(E')D_x]\bigr\}.
\end{aligned}
\label{eq:j_trace}
\end{equation}
Both spectral densities use the bare Hamiltonian, with the chosen occupations entering through $\rho_{\rm occ}$. In the atomic basis, the trace contains closed products with two current insertions. In the band basis, the energy integrals give products of interband current matrix elements divided by squared energy differences as $\eta\to0^+$. Taking this limit after summing the complete response and cancelling the valley-paired valence contribution gives Eq.~\eqref{eq:berry} for the added carriers.

Related real-space formulations express the Chern number of an insulator through spectral projectors~\cite{Kitaev2006,Bianco2011}; here we retain the occupation weights of a partially filled metallic band.

\newsec{Surface projection and reduced models} --- The surface correction is $Vg_Q(E)V^\dagger$, with $V=H_{PQ}$ and $g_Q(E)=(E-H_{QQ})^{-1}$. We first evaluate the virtual transfers at $E=U_\ell=0$, as in the illustrative Hamiltonians in the main text. Writing $a_0=t_0f$ and $a_3=t_3f^*$, the $01$ transfer is $h_{01}=-a_0^4/t_1^3$.

Adding $t_2$ or $t_3$ separately gives
\begin{equation}
\delta H_s=\begin{pmatrix}0&\delta h_s\\\delta h_s^*&0\end{pmatrix},\qquad s=2,3,
\label{eq:surface_remote_corrections}
\end{equation}
with
\begin{equation}
\delta h_2=-\frac{2a_0t_2}{t_1},\qquad
\delta h_3=\frac{3a_0^2a_3}{t_1^2}-\frac{a_3^2}{t_1}.
\label{eq:surface_remote_transfers}
\end{equation}
The two $t_2$ routes are $A_1\to B_3\to A_4\to B_4$ and $A_1\to B_1\to A_2\to B_4$. Each contains one $t_2$, one vertical dimer transfer and one $t_0$. For $t_3$, one replacement of a $t_0,t_1,t_0$ segment has three possible positions, while two replacements have one. Substituting $t_0f\simeq-\hbar v_0\pi_\tau$ and $t_3f^*\simeq-\hbar v_3\pi_\tau^*$ gives Eq.~\eqref{eq:h013}.

For $t_4$, the top surface couples to $(B_1,A_2)$ through $(t_0f,t_4f)$. The isolated-dimer propagator is $-\sigma_x/t_1$, so the leading correction, linear in $t_4$, is
\begin{equation}
\delta H_4=d_4 I,\qquad d_4=-\frac{2t_0t_4}{t_1}|f|^2.
\label{eq:surface_t4_dimer}
\end{equation}
The two terms enter and leave the dimer in opposite orders; the bottom surface gives the same result. For the linear layer-potential profile $U_\ell=(\Delta,\Delta/3,-\Delta/3,-\Delta)$ used in the numerical calculations, the normalized $01$ surface Hamiltonian also has $m_{01}(q)=\Delta[1-\tfrac23\lambda_q^2+O(\lambda_q^4)]$ to first order in $\Delta$, with $\lambda_q=\hbar v_0q/t_1$. The positive $d_4$ opposes this downward band bending.

Figure~\ref{fig:model_parameter_dependence} compares the reduced models as density and layer potential are varied, with $U_\ell=(U/2,U/6,-U/6,-U/2)$ and each model filled to the same carrier density. The $014$ error reaches $6.5\%$ on the sampled density cut and about $12\%$ near $U=108$ meV on the refined layer-potential cut. Near $n_c=10^{12}$ cm$^{-2}$ at $U=50$ meV, $0123$ gives closer agreement. Thus, the accuracy of a reduced model depends on the parameter regime.

\begin{samepage}
\newsec{Current response to a local potential} --- We calculate the response to $V(\mathbf r)=V_0e^{-r^2/(2\xi^2)}$, the same on all four layers, to first order at fixed valley polarization. The potential is screened; no self-consistent Hartree or exchange adjustment is made.\par
\end{samepage}

For eigenstates $|n\rangle$ of the unperturbed Hamiltonian, with energies $\epsilon_n$ and occupations $f_n$, the change of the density matrix is
\begin{equation}
\Delta F_{nm}=\frac{f_n-f_m}{\epsilon_n-\epsilon_m}V_{nm},\qquad V_{nm}=\langle n|V|m\rangle.
\label{eq:defect_R}
\end{equation}
The indices include band and momentum. At equal energies within the selected channel, the ratio is $f'(\epsilon_n)$. The smooth potential conserves valley. All eight bands enter, and the paired-valley valence contribution to the charge current cancels.

Transforming to atomic sites gives $\Delta F_{ij}=\sum_{nm}\psi_n(i)\Delta F_{nm}\psi_m^*(j)$ and the bond current
\begin{equation}
\Delta I_{i\to j}=\frac{2e}{\hbar}\operatorname{Im}[H_{ij}\Delta F_{ji}].
\label{eq:defect_bond}
\end{equation}
Summing over physical bonds gives the sheet current $\mathbf j(\mathbf r)$. For a slowly varying potential, $\mu\to\mu-V(\mathbf r)$ gives the local result~\cite{Cooper1997,XiaoYao2006}
\begin{equation}
\mathbf j(\mathbf r)=\frac{\partial M_z}{\partial\mu}\,\hat{\mathbf z}\times\nabla V(\mathbf r),
\label{eq:defect_limit}
\end{equation}
including both orbital-moment and Berry-curvature contributions to $\partial M_z/\partial\mu$, evaluated separately in the Supplemental Material~\cite{SM}.

For a thin sheet, the magnetic field at height $z>0$ is
\begin{equation}
B_z(\mathbf q,z)=\frac{i\mu_0}{2q}e^{-qz}[q_xj_y(\mathbf q)-q_yj_x(\mathbf q)],\qquad q=|\mathbf q|.
\label{eq:defect_field}
\end{equation}
We Fourier transform with $e^{i\mathbf q\cdot\mathbf r}$ to obtain Fig.~\ref{fig:defect_main}.

\endgroup

\onecolumngrid
\clearpage
\setcounter{equation}{0}
\setcounter{figure}{0}
\setcounter{table}{0}
\setcounter{section}{0}
\setcounter{subsection}{0}
\renewcommand{\theHequation}{supp.\arabic{equation}}
\renewcommand{\theHfigure}{supp.\arabic{figure}}
\renewcommand{\theHtable}{supp.\arabic{table}}
\renewcommand{\theHsection}{supp.\arabic{section}}
\renewcommand{\theHsubsection}{supp.\arabic{section}.\arabic{subsection}}
\phantomsection\label{sec:supplement}
\pdfbookmark[0]{Supplemental Material}{supplement}
\raggedbottom
\clubpenalty=10000
\widowpenalty=10000
\brokenpenalty=10000
\begin{center}
{\large\textbf{Supplemental Material for}}\\[4pt]
\textbf{Anomalous Hall Response and Real-Space Current Correlations in Rhombohedral Graphene}\\[4pt]
Zhen Bi and Chandra M. Varma
\end{center}
\setcounter{secnumdepth}{2}
\renewcommand{\theequation}{S\arabic{equation}}
\renewcommand{\thefigure}{S\arabic{figure}}
\renewcommand{\thetable}{S\arabic{table}}

\section{Hall response of the surface states}
We test the surface description by expanding the interior propagator with all hopping amplitudes retained. Table~\ref{tab:surface_accuracy} compares the Hall conductivity at successive orders with the full eight-band result, using the parameters of the main text. The second-order result agrees within $0.44\%$.

Let $P$ project onto the surface orbitals $(A_1,B_4)$ and $Q=1-P$ onto the remaining six orbitals, with $V=H_{PQ}$. Eliminating the interior gives~\cite{Lowdin1951}
\begin{equation}
H_{\rm eff}(E)=H_{PP}+Vg_Q(E)V^\dagger,\qquad g_Q(E)=(E-H_{QQ})^{-1}.
\label{sm:B1}
\end{equation}
The second term describes propagation from a surface orbital through the interior and back to either surface. We write $H_{QQ}=H_d+R$, where $H_d$ contains the layer potentials and the three vertical $t_1$ dimers. Treating each dimer exactly, we expand in the remaining interior couplings $R$,
\begin{equation}
g_Q^{(N)}(E,\mathbf k)=\sum_{r=0}^{N}[g_d(E)R(\mathbf k)]^r g_d(E),\qquad g_d(E)=(E-H_d)^{-1}.
\label{sm:B2}
\end{equation}
The $r=0$ term contains only propagation within isolated dimers; each insertion of $R$ connects different dimers. This expansion differs from the hopping-path expansion illustrated in Fig.~\ref{fig:surface_series}. Here $N$ counts insertions of $R$, with propagation inside each dimer already summed to all orders. The surface--interior couplings $V$ are retained in full.

The layer potentials and energy dependence of the propagator are retained in this calculation. At each momentum, we solve $H_{\rm eff}^{(N)}(E_N)p_N=E_Np_N$ for the energy $E_N$ and the two surface amplitudes $p_N$. Since the effective Hamiltonian depends on energy, its eigenvalue must equal the energy used to evaluate the propagator.

We then recover the six interior amplitudes from the Schr\"odinger equation $(E-H_{QQ})\chi=V^\dagger p$, using $g_Q^{(N)}$ for the inverse. The column vector $w_N$ collects the two surface amplitudes followed by the six interior amplitudes, in the order $(A_1,B_4;B_1,A_2,B_2,A_3,B_3,A_4)$. This gives the unnormalized state $w_N$ and the normalized state $u_N$,
\begin{equation}
w_N=\begin{pmatrix}p_N\\g_Q^{(N)}(E_N)V^\dagger p_N\end{pmatrix},\qquad u_N=\frac{w_N}{\sqrt{w_N^\dagger w_N}}.
\label{sm:B3}
\end{equation}
The interior amplitudes vary with momentum and contribute to the Berry curvature even when their weight is small. We therefore calculate the Hall conductivity from the normalized eight-orbital state $u_N$. At each order, the chemical potential is adjusted to keep the carrier density fixed. 

\begin{table}[htbp]
\caption{Hall conductivity of the reconstructed surface states at $T=0$. Relative errors refer to the full eight-band result.}
\label{tab:surface_accuracy}
\begin{ruledtabular}
\begin{tabular}{lrr}
Approximation & $\sigma_{xy}/(e^2/h)$ & Relative error\\
First order ($N=1$) & $-0.155352$ & $32.7\%$\\
Second order ($N=2$) & $-0.229878$ & $0.44\%$\\
Fourth order ($N=4$) & $-0.230906$ & $0.0074\%$\\
Full eight-band model & $-0.230888$ & ---\\
\end{tabular}
\end{ruledtabular}
\end{table}

\section{Current induced by a smooth potential}
For Fig.~\ref{fig:defect_main}, we calculate the response of the full eight-band model at $T=0$ to the screened potential $V(\mathbf r)=V_0e^{-r^2/(2\xi^2)}$, with $V_0=0.1$ meV and $\xi=100$ nm. The potential acts equally on all orbitals. We work to first order in $V$ at fixed spin and valley polarization, without a self-consistent Hartree or exchange adjustment.

The single-particle density matrix is $F_{ij}=\langle c_j^\dagger c_i\rangle$, where $i,j$ label atomic sites. Its diagonal entries give the occupations and its off-diagonal entries give the intersite coherence that enters the bond current. The potential mixes the unperturbed eigenstates, producing the change $\Delta F$ in Eq.~(\ref{eq:defect_R}).

A Fourier component $V(\mathbf q)$ connects states at $\mathbf k_-=\mathbf k-\mathbf q/2$ and $\mathbf k_+=\mathbf k+\mathbf q/2$, with band eigenvectors $u_{n\pm}$ and energies $\epsilon_{n\pm}$. Since $\xi$ is much larger than the lattice spacing, scattering between valleys is negligible. The potential is proportional to the identity in orbital space, so its matrix element is $V(\mathbf q)\langle u_{n+}|u_{m-}\rangle$.

The filled valence bands give cancelling charge-current contributions from the two valleys. We therefore use the added-carrier occupations $f_{n\pm}$, which are nonzero only in the conduction band of the selected spin and valley. All eight bands enter the response through mixing with these occupied states. Transforming Eq.~(\ref{eq:defect_R}) to the orbital basis gives
\begin{equation}
\frac{\Delta F_{\alpha\beta}(\mathbf k_+,\mathbf k_-)}{V(\mathbf q)}
=\sum_{nm}\frac{f_{n+}-f_{m-}}{\epsilon_{n+}-\epsilon_{m-}}u_{n+,\alpha}\langle u_{n+}|u_{m-}\rangle u_{m-,\beta}^*.
\label{sm:defect_matrix_response}
\end{equation}
Here $n,m$ label bands, $\alpha,\beta$ label orbitals, and $\Delta F_{\alpha\beta}=\Delta\langle c^\dagger_{\mathbf k_-,\beta}c_{\mathbf k_+,\alpha}\rangle_{\rm add}$. The occupation difference removes contributions from mixing equally occupied states, while the eigenvector components convert the band response to orbital matrix elements. For equal energies in the same occupation channel, the quotient is $f'(\epsilon)$, giving $-\delta(\epsilon-\mu)$ in the partially filled band at $T=0$.

Fourier transforming $\Delta F$ to atomic sites gives $\Delta I_{i\to j}=(2e/\hbar)\operatorname{Im}[H_{ij}\Delta F_{ji}]$. Distributing each current along its bond yields the sheet current
\begin{equation}
\mathbf j(\mathbf q)=\frac1{A_{\rm cell}}\sum_b\mathbf d_b\operatorname{sinc}(\mathbf q\cdot\mathbf d_b/2)\Delta I_b(\mathbf q),
\label{eq:defect_sheet}
\end{equation}
where $\Delta I_b$ is referred to the bond midpoint, $\mathbf d_b$ is its in-plane displacement, $A_{\rm cell}=\sqrt3a^2/2$, and $\operatorname{sinc}x=\sin x/x$. The sinc factor comes from distributing the current along the bond. We obtain the orbital magnetic field from Eq.~(\ref{eq:defect_field}); the factor $e^{-qz}$ suppresses short-wavelength components at larger heights.

For a slowly varying potential, each region can be treated locally as a uniform system at chemical potential $\mu-V(\mathbf r)$. The change in orbital magnetization is then $\delta M_z=-(\partial M_z/\partial\mu)V$, which gives $\mathbf j=(\partial M_z/\partial\mu)\hat{\mathbf z}\times\nabla V$. This is the local approximation in Fig.~\ref{fig:defect_main}, evaluated with the same eight-band model. Differentiating the orbital magnetization~\cite{XiaoShiNiu2005,Thonhauser2005,Ceresoli2006,Xiao2010} gives
\begin{equation}
\frac{\partial M_z}{\partial\mu}=\sum_n\int\frac{\mathrm d^2k}{(2\pi)^2}\left[m_{nz}\delta(\epsilon_n-\mu)+\frac{e}{\hbar}\Omega_{nz}\Theta(\mu-\epsilon_n)\right].
\label{sm:magnetization_derivative}
\end{equation}
Here $m_{nz}$ and $\Omega_{nz}$ are the orbital magnetic moment and Berry curvature. The first term is the contribution of states added at the Fermi surface and gives $44.382$ nA/meV. The second follows from the explicit chemical-potential dependence of the Berry-curvature term in the magnetization and gives $8.945$ nA/meV. This term equals $-\sigma_{xy}/e$ and accounts for only $16.8\%$ of the total. The Hall conductivity therefore fixes only part of the current response.

For the chosen potential, the microscopic and local current profiles differ by at most $0.50\%$ of the peak local current, with both calculated to first order in $V$. We also checked the long-wavelength response against Eq.~\eqref{sm:magnetization_derivative} and the magnetic field against a real-space Biot--Savart calculation.

\end{document}